\documentclass[twocolumn,showpacs,showkeys,preprintnumbers,amsmath,amssymb]{revtex4}  

\usepackage{graphicx}
\usepackage{dcolumn}
\usepackage{bm}
\usepackage{multirow}
\usepackage{color}
\usepackage{hyperref}

\begin{document}

\title{Strain tuning of optical emission energy and polarization\\ in monolayer and bilayer MoS$_2$}

\author{G.~Wang$^1$}
\author{C.R.~Zhu$^1$}
\author{B.L.~Liu$^1$}
\author{X.~Marie$^{2}$}
\author{Q.X.~Feng$^3$}
\author{X.X.~Wu$^1$}
\author{H.~Fan$^1$}
\author{P.H.~Tan$^3$}
\author{T.~Amand$^2$}
\author{B.~Urbaszek$^2$}

\affiliation{%
$^1$Beijing National Laboratory for Condensed Matter Physics, Institute of Physics, Chinese Academy of Sciences, Beijing 100190, China}
\affiliation{%
$^2$Universit\'e de Toulouse, INSA-CNRS-UPS, LPCNO, 135 Av. de Rangueil, 31077 Toulouse, France}
\affiliation{%
$^3$State Key Laboratory of Superlattices and Microstructures, Institute of Semiconductors, Chinese Academy of Sciences, Beijing,100083, China}

\date{\today}

\begin{abstract}
We use micro-Raman and photoluminescence (PL) spectroscopy at 300K to investigate the influence of uniaxial tensile strain on the vibrational and optoelectronic properties of monolayer and bilayer MoS$_2$ on a flexible substrate. The initially degenerate $E^1_{2g}$ Raman mode is split into a doublet as a direct consequence of the strain applied to MoS$_2$ through Van der Waals coupling at the sample-substrate interface. We observe a strong shift of the direct band gap of 48meV/(\% of strain) for the monolayer and 46meV/\% for the bilayer, whose indirect gap shifts by 86meV/\%. We find a strong decrease of the PL polarization linked to optical valley initialization for both monolayer and bilayer samples, indicating that scattering to the spin-degenerate $\Gamma$ valley plays a key role.

\end{abstract}

\pacs{78.60.Lc,78.66.Li}
                           \keywords{valley selectivity, monolayer MoS2, Photoluminescence }
\maketitle

\begin{figure}
\includegraphics[width=0.47\textwidth]{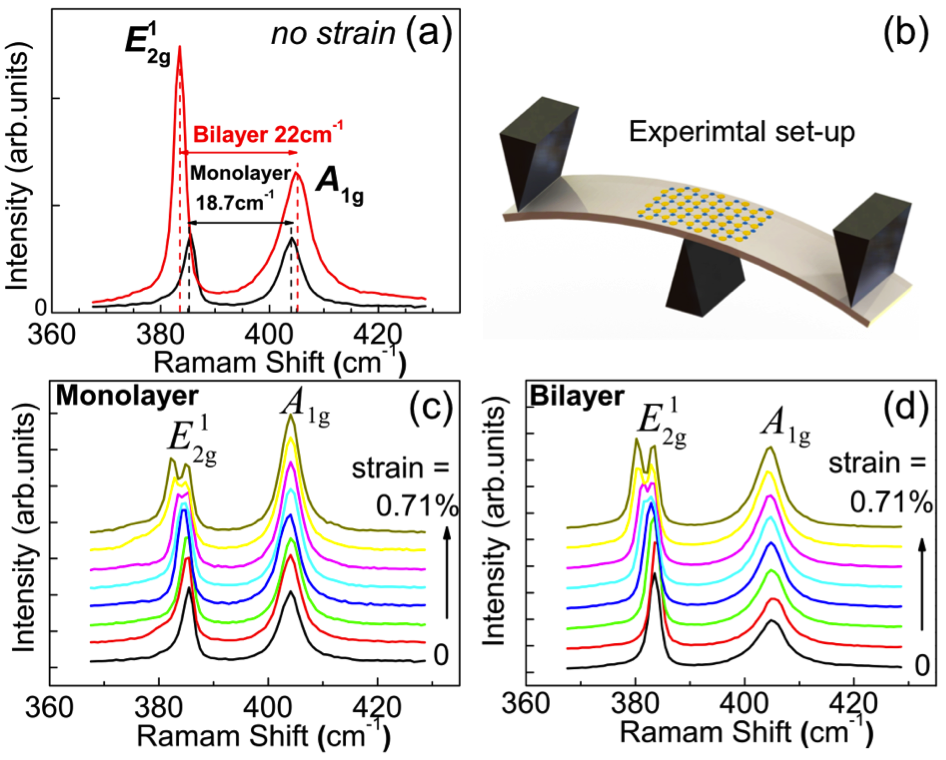}
\caption{\label{fig:fig1} (a) The monolayer (black curve) and bilayer (red) $MoS_2$ regions are identified by Raman spectroscopy. (b) The 3 point bending apparatus. Note that the distance between the two top contact points is 6~cm  whereas the extension of a Mo$S_2$ flake is in the $\mu m$ range. (c) The $E^1_{2g}$ Raman mode splits into two modes as tensile strain is applied to the monolayer sample. (d) same as (c) but for bilayer sample.}
\end{figure} 
\textit{Introduction.---}
Transition metal dichalcogenides such as MoS$_2$ emerge as an exciting class of atomically flat, two-dimensional materials for electronics and optoelectronics. In contrast to graphene, monolayer MoS$_2$ has a direct bandgap \cite{Mak:2010a,Splendiani:2010a} in the visible region of the optical spectrum and has been used as the active region of field effect transistors \cite{Radisavljevic:2011a}, complex electronic circuits \cite{Wang:2012a} and light emitting diodes \cite{Sundaram:2013a}. Another important difference to graphene is the broken inversion symmetry of monolayer MoS$_2$, which can be directly used for second harmonic generation in non-linear optics \cite{Kumar:2013a}. The combined presence of inversion symmetry breaking and strong spin orbit coupling \cite{Zhu:2011a} allows simple optical k-valley initialisation with circularly polarized lasers \cite{Cao:2012a,Mak:2012a,Zeng:2012a,Sallen:2012a}. This opens up very exciting possibilities of manipulating carriers in valleys with contrasting Berry phase curvatures \cite{Xiao:2010a}, which allows in principle the observation of the Valley hall effect \cite{Xiao:2012a}. The valley and spin properties are closely linked to the crystal symmetry and are expected to be modified through the application of mechanical strain. The role of strain is also important for practical devices using MoS$_2$ on flexible substrates. In unstrained samples the indirect bandgap of monolayer MoS$_2$ is just above the direct bandgap \cite{Splendiani:2010a,Zhu:2011a,Cheiwchanchamnangij:2012a}. Strain will therefore have important consequences as the indirect transition approaches the direct transition energy before monolayer MoS$_2$ becomes indirect for strain exceeding 1.5\% \cite{Shi:2013a}.\\
\indent Here we apply relatively small tensile uniaxial strain of up to 0.8\% to observe striking changes in the Raman spectra of monolayers and bilayers. Drastic changes in optical emission properties are observed: the bandgap is shifted by several tens of meV and the PL polarization changes by 40\% for the monolayer and $100\%$ for the bilayer (i.e. the finite PL polarization is tuned to zero). These results suggest that scattering to the spin-degenerate $\Gamma$ valley \cite{Zhu:2011a} plays a key role as it becomes more efficient as strain increases.\\
\indent \textit{Samples and Set-up.---} 
In order to controllably induce strain, the MoS$_2$ flakes are obtained by micro-mechanical exfoliation of natural bulk MoS$_2$ crystals (SPI supplies, USA) onto a flexible substrate, which is a polyethylene terephthalate (PET) film (9cm long, 1cm wide, 1mm thick). The MoS$_2$ flakes remain at a fixed position on the substrate due to van der Waals attraction. Uniaxial strain is applied to MoS$_2$ through bending the PET film in a three-point apparatus (distance between two extreme points = 6cm). To achieve maximum strain, the MoS$_2$ flake is positioned exactly on top of the center point, see Fig. \ref{fig:fig1}(b). The induced strain $\epsilon$ is given by $\epsilon=t/2R$, where $t=1mm$ is the thickness of PET film and $R$ is the radius of curvature of the bent substrate. $R$ is evaluated by measuring the displacement of the central point of the substrate (i.e. where the strain is applied). The angle at which the strain is applied with respect to the MoS$_2$ in-plane crystal orientation can be adjusted as the bending apparatus is mounted on a rotation stage. Raman spectroscopy is performed in a back-scattering geometry using a Jobin-Yvon HR800 Raman system equipped with a liquid-nitrogen cooled CCD and laser excitation at 2.6eV. This allows to verify that the MoS$_2$ flakes adhere well to the bent substrate during the measurements (i.e. no sample slippage occurs).  Micro-photoluminescence (PL) spectra are measured with a 100x long-working-distance objectives using a He-Ne laser at 1.95~eV for excitation. The laser beam is passed through a Soleil Babinet Compensator to create circularly-polarized light. The PL polarization $P_c$ defined as $P_c = \frac{I_{\sigma+}-I_{\sigma-}}{I_{\sigma+}+I_{\sigma-}}$ is analyzed by a quarter-wave plate placed in front of a Glan-Thomson linear polarizer. Here $I_{\sigma+}(I_{\sigma-})$ denotes the intensity of the $\sigma^+(\sigma^-)$ polarized emission. All experiments are carried out at room temperature.\\
\indent \textit{Experimental Results.---} Before applying strain, the monolayer and bilayer regions are localised by micro-Raman spectroscopy \cite{Lee:2010a,Korn:2011a} as the $E^1_{2g}$ and the $A_{1g}$ modes are identified \cite{Wieting:1971a}, see Fig.\ref{fig:fig1}(a). As the tensile strain is applied to the monolayer sample, we observe a clear splitting of the degenerate $E^1_{2g}$ mode into two distinct peaks in Fig.\ref{fig:fig1}(c). In a recent report only a broadening, not a splitting of the $E^1_{2g}$ mode has been observed, although a splitting was predicted by the same authors \cite{Rice:2013a}. Observing a clear splitting here for the monolayer and also the bilayer sample (see Fig.\ref{fig:fig1}(d)) is most likely due to a higher spectral resolution and a more homogeneous strain distribution in our sample.  The results of Fig.\ref{fig:fig1}(c) and (d) confirm strain is successfully applied to the sample via Van der Waals force at the substrate-sample interface. The applied strain is limited to 0.8\% to avoid sample slippage \cite{elastic}. 

\begin{figure}
\includegraphics[width=0.5\textwidth]{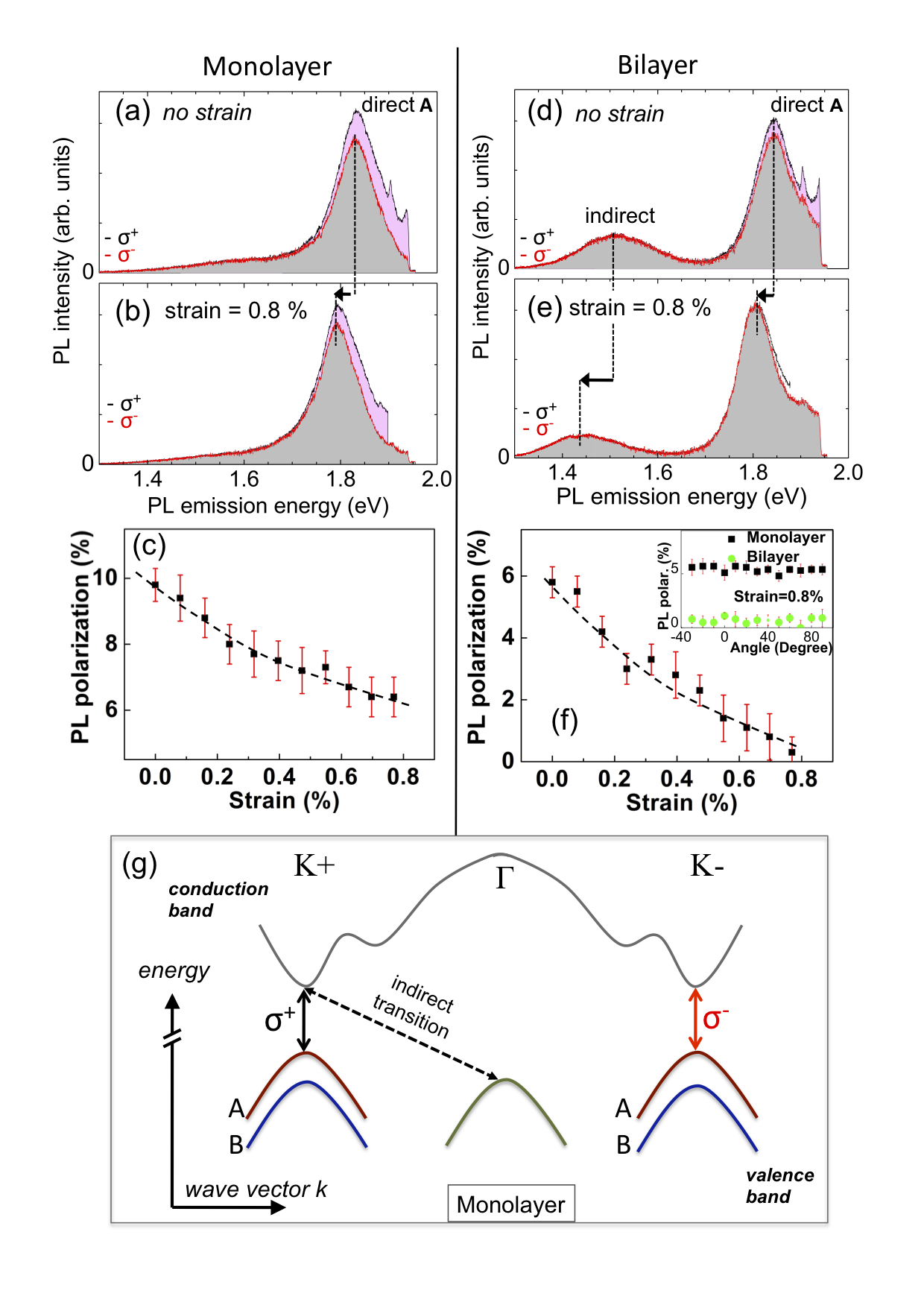}
\caption{\label{fig:fig2} Laser excitation at 1.95eV with $\sigma^+$ polarization. (a) PL emission detected for $\sigma^+$ and $\sigma^-$ polarization for MoS$_2$ \textbf{monolayer} at zero strain (b) as (a) but for applied strain of 0.8\%. (c) Circular polarization degree of the PL as a function of the applied tensile strain. Dotted lines guide to the eye (d) PL emission detected for $\sigma^+$ and $\sigma^-$ polarization for MoS$_2$ \textbf{bilayer} at zero strain (e) as (d) but for applied strain of 0.8\%. (f) PL Circular polarization degree as a function of the applied tensile strain. Dotted lines are guides to the eye. Inset: The applied strain is kept constant at 0.8\% and the PL polarisation is measured as a function of the angle of the applied strain with respect to the in-plane mono- and bilayer crystal orientation. (g) Scheme of the band structure for monolayer MoS$_2$ at zero strain. K and $\Gamma$ valleys and the associated interband transitions are marked, the conduction band spin splitting in the meV range \cite{Zhu:2011a} is not shown.
}
\end{figure} 

\begin{figure}
\includegraphics[width=0.47\textwidth]{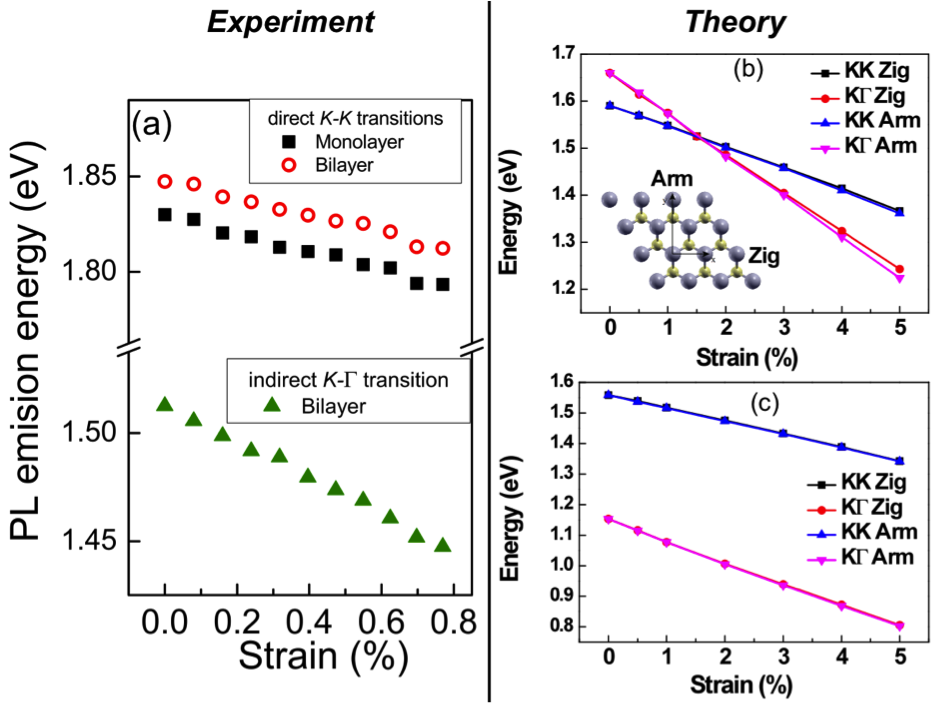}
\caption{\label{fig:fig3} (a) Measured energy shift for the direct recombination of MoS$_2$ monolayer (squares) bilayer sample (circles), and indirect recombination for bilayer (triangles) for applied strain between 0 and 0.8\%. (b) calculation (see main text) of the evolution of the direct (KK) and indirect (K$\Gamma$) bandgap for a MoS$_2$ monolayer with strain from 0 to 5\% applied along the zig-zag and armchair directions.  (c) Same as (b) but for a MoS$_2$ bilayer.
}
\end{figure}

In Fig.\ref{fig:fig2}(a) we present the PL emission of a typical monolayer sample excited with a $\sigma^+$ circularly polarized laser. We detect a circular polarization degree in the order of 10\% due to the chiral optical selection rules in MoS$_2$ monolayers \cite{Cao:2012a,Xiao:2012a} linked to K-valley polarization \cite{Mak:2012a,Zeng:2012a,Sallen:2012a,Kioseoglou:2012a}. As the strain is increased up to 0.8\%, we observe two striking changes in Fig.\ref{fig:fig2}(b): First, the emission is shifted by several tens of meV to lower energy. Second, the PL polarization degree $P_c$ decreases. This is confirmed in Fig.\ref{fig:fig2}(c) where we measure a systematic decrease of $P_c$ with increasing strain.\\
\indent Also for the bilayer sample we observe a shift of the PL emission energy associated with the direct transition, compare Fig.\ref{fig:fig2}(d) and (e). This emission is polarized in the absence of strain to about 6\%. Note that the PL polarization of the indirect optical transitions at lower energy is zero, as expected (see also discussion below). Under the maximum tensile strain of 0.8\% the PL polarization of both transitions (direct and indirect) is zero and they are shifted to lower energy in Fig.\ref{fig:fig2}(e). We observe that $P_c$ for the bilayer can be tuned continuously from 6\% to zero as strain is increased in Fig.\ref{fig:fig2}(f).
By rotating the sample we have verified that the decrease in polarization does not depend on the direction of the tensile strain applied in the plane, see inset of Fig.\ref{fig:fig2}(f) \cite{strainiso} in good agreement with our calculastions presented in Fig.\ref{fig:fig3}(b) and (c).

The systematically measured shift in PL emission energy due to a strain induced reduction in bandgap is plotted in Fig.\ref{fig:fig3}(a). We observe a strong shift of the \textit{direct} band gap of 48meV/\% of strain for the monolayer and 46meV/\% for the bilayer. The shift of the \textit{indirect} gap of the bilayer of 86meV/\% is nearly twice as high. These shifts are in agreement with our calculations and the calculations reported in the literature, see for example \cite{Feng:2012a,Shi:2013a,Yun:2012a,strainexciton}. Our density functional theory (DFT) calculations employ the projector augmented wave (PAW) method encoded in Vienna \emph{ab initio} simulation package(VASP) \cite{Kresse:1993a,Kresse:1996a,Kresse:1996b}, and the Perdew-Burke-Ernzerhof (PBE) exchange correlation functional \cite{Perdew:1996a} is used. Throughout this work, the cutoff energy of 400 eV is taken for expanding the wave functions into plane-wave basis. For the relaxation of bilayer MoS$_2$, the long-range van der Waals interactions is accounted by means of a semi empirical DFT-D2 approach proposed by \textcite{Grimme:2006a}. The calculated parameters for monolayer and bilayer MoS$_2$ are a=3.189 \AA ~and a=3.196 \AA ~with $15\times15\times1$ $\Gamma$ centered k-point grids.  A 20 \AA ~vacuum layer is used in our calculation to avoid interactions between slabs. The convergence for energy is chosen as $10^{-5}$ eV between two steps and the maximum Hellmann-Feynman force acting on each atom is less than 0.01 eV/\AA ~upon ionic relaxation. Fig.\ref{fig:fig3}(b) shows the evolution of the calculated band gap with strain using PBE functional. Although the bandgap is underestimated, the calculated shift of direct bandgap is 42 meV/\% for the monlayer and bilayer and the shift of indirect bandgap is 76 meV/\% for the bilayer, which is good agreement with our experimental observation. Our calculations expect a crossing of direct and indirect bandgap at $\epsilon= 1.5\%$, in agreement with the prediction of \textcite{Shi:2013a}, which we can not verify directly in our experiment as the applied strain is limited to $\epsilon = 0.8\%$.\\
\indent \textit{Discussion.---} 
For a relatively small uniaxial strain amplitude we observe profound changes in the electronic structure of monolayer and bilayer MoS$_2$. We first analyse the observations for the monolayer sample.\\
\indent \textbf{Monolayer} MoS$_2$ has a direct bandgap at the K-point with chiral optical selection rules \cite{Xiao:2012a,Cao:2012a}, see Fig.\ref{fig:fig2}(g). When a laser is resonant with the \textbf{A}-valence band state  ($K^{\pm}$ valley) to conduction band transition, $\sigma^+$ polarized light will result in the creation of a conduction electron in the $K^+_c$ valley, $\sigma^-$ polarized light creates a $K^-_c$ electron. In emission the same selection rules apply, so in the absence of intervalley carrier transfer (and spin flips) the emission is expected to be strongly polarized, as is observed at low temperature \cite{Cao:2012a,Mak:2012a,Zeng:2012a,Sallen:2012a}. The valley and spin states are less stable and less well defined at room temperature and lower PL polarization degrees are observed as also reported here. In our experiments the exciting photon energy is close to the \textbf{B} K-valence to conduction band transition. As a result both \textbf{B} and \textbf{A} bands are excited due to energy broadening by impurities, phonons and substrate imperfections \cite{Song:2013a,Kioseoglou:2012a}. The B-excitons have to relax in energy, which can lead to a change in valley if high k-value phonons are emitted. \\
\indent It is important to note that although the fundamental gap of monolayer MoS$_2$ is direct, the indirect gap between the valence band maximum $\Gamma_v$ and the degenerate conduction band minima $K^+_c$ and $K^-_c$ is very close in energy \cite{Cheiwchanchamnangij:2012a} as indicated in Fig.\ref{fig:fig2}(g). This means at room temperature, as energy levels are broadened and due to high phonon occupation numbers, the $\Gamma_v$ states will play an important role in optics and transport, as has been theoretically predicted \cite{Song:2013a,Kormanyos:2013a}. Although the phonon-assisted indirect absorption is a second order process, it cannot be neglected since it has many more available final states compared with the direct absorption \cite{Elliot:1957a}. The chiral optical selection rules that allow optical valley initialisation do not apply to the $\Gamma_v \leftrightarrow K^{\pm}_c$ transitions as the large spin splitting of 150 meV in the $K^{\pm}_v$ valleys is absent at the $\Gamma$ point where the valence states are degenerate \cite{Zhu:2011a,Cheiwchanchamnangij:2012a}. Therefore phonon assisted light absorption and emission involving the $\Gamma$ valence states will be essentially unpolarized.\\
\indent As \textbf{strain} is applied to a \textbf{monolayer} sample, theory predicts that the $\Gamma_v \leftrightarrow K^{\pm}_c$ transition becomes the fundamental transition for $\epsilon > 1.5\%$, see our calculation in Fig.\ref{fig:fig3}(b). So the impact of the indirect transitions on the optical properties will be the more important the higher the strain is. We propose that the scattering to the spin-degenerate $\Gamma$ valley is at the origin of the clear decrease of the PL polarization of the \textbf{A}-transition as a function of the applied strain (Fig.\ref{fig:fig2}(c)). First, hole scattering from the K to the $\Gamma$ valley can occur after absorption at the direct optical bandgap. Second, the absorption of a photon and a phonon can result in an indirect transition from the valence to the conduction band \cite{Elliot:1957a}. Both scenarios will result in a global decrease of the PL polarization. As the strain is increased, this effect is exalted in our experiment as we do not move our laser energy which is constant at 1.95eV. In general, the more off-resonance the optical excitation, the lower the polarization on the ground state due to losses during energy relaxation. Although other physical properties such as the phonon dispersion also change with strain, we feel that the drastic change in transitions energy and the associated competition between direct and indirect transitions are the main origin for the observed change in PL polarisation.  \\
\indent We now discuss the observed lowering of the PL polarization of the \textbf{bilayer} with the applied strain. As for the monolayer, the valence states in the K-valley (around the valence band maximum) are split into a high energy \textbf{A} and a low energy \textbf{B} band. But contrary to the monolayer case, both bands are spin-degenerate, so for an ideal bilayer, there is no spin splitting at the $K^{\pm}$ points in the valence band as the crystal posses inversion symmetry \cite{Xiao:2012a,Cao:2012a}. For zero external strain we observe a 6\% PL polarization, so a spin splitting at the K-point induced by symmetry breaking due to surface and interface effects is likely (assuming fast electron and hole spin relaxation compared to the radiative lifetime).
A non-zero bilayer PL polarization has been reported before \cite{Mak:2012a,Wu:2012a}. Inversion symmetry breaking also manifests itself by second harmonic generation \cite{Kumar:2013a}, forbidden for a perfectly symmetrical bilayer. The bilayer emission is very rich in information due to the co-existence of indirect and direct emission: at zero strain the direct optical emission is circularly polarized, whereas the indirect emission (unfortunately not detectable for the monolayer) is unpolarized in Fig.\ref{fig:fig2}(d). It is reasonable to assume that the splitting induced between the A and B valence band at the K-point is smaller than the value of 150meV for the monolayer. So qualitatively the same arguments used to explain the decrease in polarization for the monolayer sample apply to the bilayer sample. The difference in behaviour is quantitative, as the relative change in $P_c$ in the bilayer (100\%) is bigger than in the monolayer (40\%). This is very likely due to the fact that the indirect transition is already the fundamental bandgap as we start the measurements at zero strain. As we keep our laser energy constant we excite more and more non-resonantly as the strain increases. Also, the small valley polarization created in the bilayer will be more fragile, as the K-point splitting is smaller. The application of strain could also restore inversion symmetry for the bilayer, which would contribute in addition to lower the PL polarization \cite{Wu:2012a}. 

\indent \textit{Conclusion.---} The unique coexistence of direct and indirect exciton transitions in uniaxially strained MoS$_2$ monolayers and bilayers has been investigated in the context of valley polarization. For monolayers, this coexistence is very promising for p-type Gunn diodes in applied electric fields \cite{Song:2013a} and has to be further investigated for realistic Valley Hall experiments. A stronger separation between K and $\Gamma$ valence bands is expected for monolayer WSe$_2$\cite{Zhu:2011a}, another promising dichalcogenide material with a direct gap also in the visible range \cite{Jones:2013a}. \\
\indent \textit{Acknowledgements.---} We acknowledge partial funding from National Basic Research Program of China (2009CB930502, 2010CB922904) and National Science Foundation of China (grant number 11174338, 10911130356(SPinman)); CAS Grant No. 2011T1J37; Labex NEXT Project "Valleyhall" and ERC St.Gr. OptoDNPcontrol (B.U.).\\
\indent \textit{Note added.---} After completing these experiments, we became aware of two preprints reporting similar results on Raman and bandgap shifts \cite{Conley:2013a,He:2013a}.


\begin{thebibliography}{37}
\bibitem[{ela()}]{elastic}
\bibinfo{note}{Our experiments are carried out in the elastic regime: For a
  given strain, we obtain the same results if the strain is varied from zero to
  0.8\% or from 0.8\% to zero.}
\bibitem[{str({\natexlab{a}})}]{strainiso}
\bibinfo{note}{As MoS$_2$ has trigonal C$_{3v}$ symmetry, it is enough to cover
  120 deg. to verify the isotropy.}
\end{thebibliography}
\end{document}